\documentclass[aps,pra,reprint,showpacs,superscriptaddress,longbibliography]{revtex4-1}
\usepackage{mathtools,amssymb,graphicx,units}
\usepackage[usenames,dvipsnames]{color}
\usepackage[plainpages=false,pdfpagelabels,colorlinks=true,linkcolor=red,urlcolor=blue,citecolor=blue,pdftitle={Title},pdfauthor={},pdfdisplaydoctitle=true,pdfduplex=DuplexFlipLongEdge]{hyperref}

\renewcommand{\vec}[1]{\boldsymbol{#1}}

\begin{document}
\title{Many-body self-localization in a translation-invariant Hamiltonian}
\author{Rubem Mondaini}
\email{rmondaini@csrc.ac.cn}
\affiliation{Beijing Computational Science Research Center, Beijing 100193, China}
\author{Zi Cai}
\email{zcai@sjtu.edu.cn}
\affiliation{Key Laboratory of Artificial Structures and Quantum
Control, Department of Physics and Astronomy, Shanghai Jiao Tong University, Shanghai 200240, China}

\begin{abstract}
We study the statistical and dynamical aspects of a translation-invariant Hamiltonian, without quench disorder, as an example of the manifestation of the phenomenon of many-body localization. This is characterized by the breakdown of thermalization and by information preservation of initial preparations at long times. To realize this, we use quasi-periodic long-range interactions, which are now achievable in high-finesse cavity experiments, to find evidence suggestive of a divergent time-scale in which charge inhomogeneities in the initial state survive asymptotically. This is reminiscent of a glassy behavior, which appears in the ground state of this system, being also present at infinite temperatures.
\end{abstract}

\maketitle

\section{Introduction}
For a quantum particle moving in a disordered medium, the coherent backscattering from randomly distributed impurities may localize all the quantum eigenstates of the system and give rise to insulating behavior. This phenomenon, known as Anderson localization~\cite{Anderson_58}, is to be distinguished from another localization mechanism driven by a strong mutual repulsion of particles instead of disorder, namely, Mott localization~\cite{Mott_49, Mott_68}. Recently, it was discovered that even in the presence of small interactions Anderson localization is robust~\cite{Fleishman_80,Altshuler_Gefen_97,Gornyi_Mirlin_05,Basko_06,Basko_08}; this was then numerically verified for a variety of quantum systems~\cite{Oganesyan_Huse_07,Znidaric_08,Monthus_10,Pal_10,Canovi_11,Khatami_Rigol_12,Bardarson_Pollmann_12,Iyer_Oganesyam_13,Serbyn_13_a,Serbyn_13_b,Serbyn_14,Huse_Nandkishore_14,Nandkishore_14,Pekker_14,Chandran_15,Luitz_15,BarLev_15,Mondaini_15,Vasseur_15,Soumya_Schomerus_15}. This quantum many-body phenomenon, dubbed many-body localization (MBL), highlights the rich physics that arises from the competition between disorder and interactions, which ultimately results in the breakdown of ergodicity and the absence of transport, even at finite energy densities, when disorder is sufficiently large.

An outstanding question is whether (many-body) localization can arise in translation-invariant systems in the absence of disorder\cite{Carleo_12,Grover_Fisher_14, DeRoeck_14,Schiulaz_14,Roeck_14,Schiulaz_15,Barbiero_Menotti_15,van_Horssen_15,Papic_15,Kim_Haah_2016,Yao_Laumman_16}.  The search of the disorder free localization can be traced back to Kagan and Maksimov's work on Helium mixtures consisting of two species of particles, light and heavy, where the inter-species interactions make heavy particles generate an effective random quasistatic potential which blocks the diffusion of light ones, thus localizing them~\cite{Kagan_Maksimov_84, Kagan_Maksimov_85}. However, recent studies suggest that such a system exhibits only transient subdiffusive dynamics, while its long-time dynamics is still ergodic, coining the term quasi-MBL to describe them~\cite{Papic_15,Yao_Laumman_16}. A recent intriguing proposal of disorder-free localization involves an exactly solvable spin-fermion model with an extensive number of conserved quantities~\cite{Smith_2017} that breaks ergodicity. In this paper, we propose a different approach to realize the disorder-free (many-body) localization: we study a uniform system composed of single-species identical particles, where interactions between a quantum particle and the others effectively serve as a spontaneously emergent disorder that in turn localizes the particle itself. The essential ingredient here is the inter-particle interaction with a peculiar long-range nature, which gives rise to a glassy behavior of the ground-state as well as glassy dynamics at finite energy densities.
                                            
Quantum many-body systems with long-range interactions in atomic, molecular, and optical systems have attracted considerable interest. Typical examples include dipole-dipole interactions between atoms or molecules with large dipolar momentum~\cite{Ni_08, Stuhler_05,Baier2015}, Van der Waals interactions between atoms in Rydberg state~\cite{Heidemann_08,Valado_16}, and variable-range interactions between ultra-cold atoms in high-finesse cavities~\cite{Baumann_10, Ritsch_13, Landig_16} or trapped ions~\cite{Islam_13,Schneider_12}. Compared to the cases of short-range interacting models, the localization phenomena are much less explored in long-range quantum lattice systems. It is well-known that the (Anderson) localization phenomenon strongly depends on the dimension of the system~\cite{Abrahams1979}, thus it is natural to expect that novel localization behavior may emerge in long-range quantum lattice systems whose dimension may be not well-defined.

\section{Model}
We consider a one-dimensional (1D) hard-core bosonic model with infinite long-range interactions whose Hamiltonian reads,
\begin{equation}
 \label{eq:hamilt}
 \hat H = -J\sum_i \left(\hat a_i^\dagger \hat a_{i+1} + h.c\right)  - \sum_{i<j}V_{ij}\left(\hat n_i -\frac{1}{2}\right)\left(\hat n_{j} -\frac{1}{2}\right),
\end{equation}
where $\hat a_i$ ($\hat a^\dagger_i$) denotes the annihilation (creation) operator for hard-core bosons, and $\hat n_i$ is the local density operator. We choose $V_{ij} = \frac{V}{L} \cos[2\pi p|i-j|]$, where $V > 0$ denotes the strength of the interactions, and $L$ is the length of the 1D lattice. The $1/L$ factor in the interaction term guarantees that the total interacting energy in the ground state linearly scales with the system size for sufficiently large $L$. The long-range interaction is periodic, $V_{ij} = V_{i+T,j}$, with $T = 1/p$ being the period of the interactions that can be either commensurate or incommensurate with the lattice constant $a_0$, which is set to unit. A key observation is that the long-range interaction is translation invariant, $V_{ij} = V_{|i-j|}$, so is the total Hamiltonian(\ref{eq:hamilt}). We work at half-filling throughout the paper.

\section{Glassy ground-state}

\begin{figure}[!tb] 
  \includegraphics[width=0.95\columnwidth]{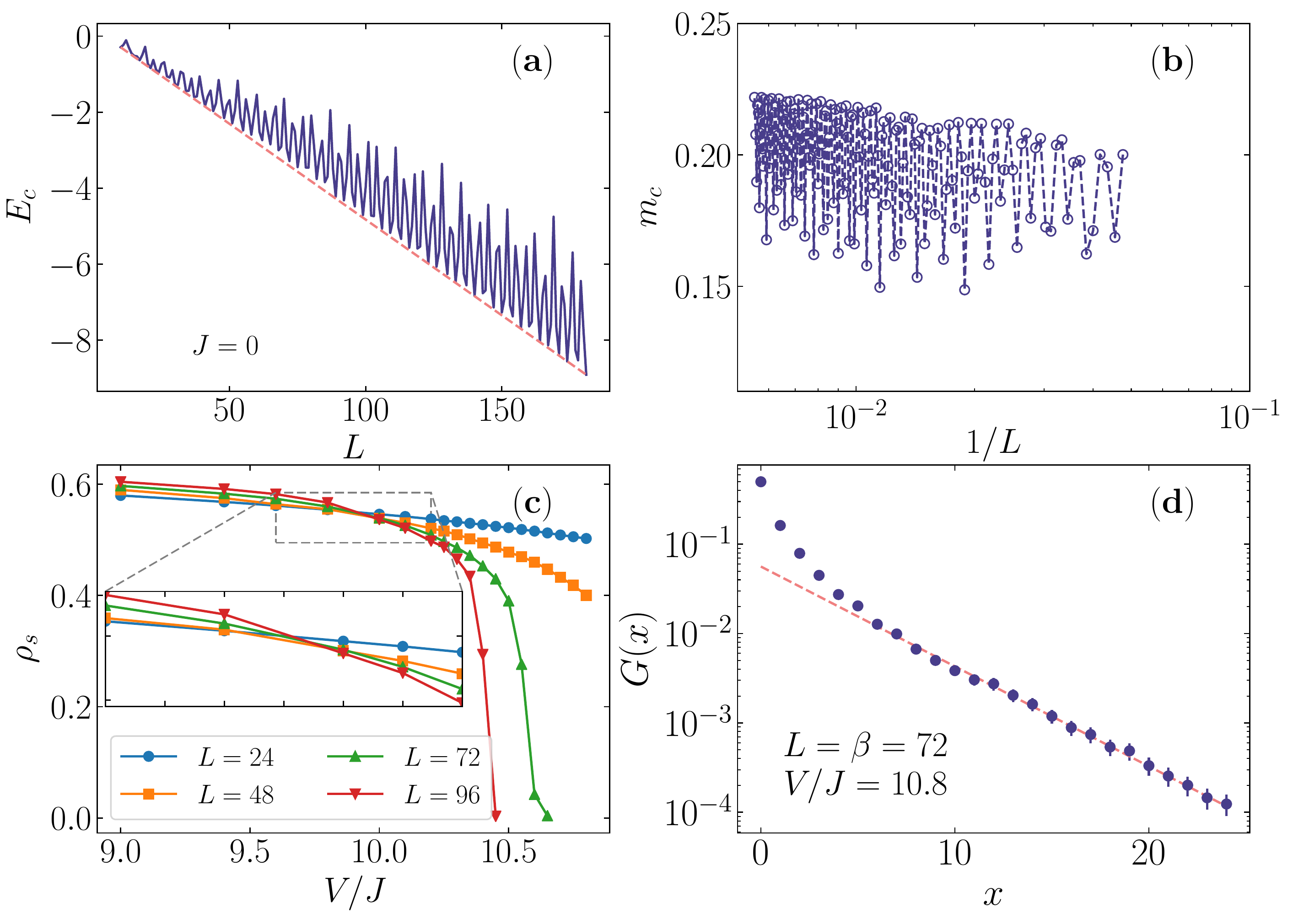}
 \vspace{-0.2cm}
 \caption{(Color online) System size dependence of the total energy (a) and order parameter of the incommensurate CDW phase (b) of the classical ground state with $J=0$. (c) Superfluid density $\rho_s$ as a function of $V$ for different lattice sizes $L$ and $\beta$ ($L/\beta = 1$); the inset is a zoom-in near the phase transition point. (d) The single-particle correlation function $G(x)$ in the strongly coupling phase for a system with $L = \beta = 72$ and $V = 10.8J$, highlighting an exponential decay (dashed line) at large separations.}
 \label{fig:mf_qmc_plots}
\end{figure}

The ground state of the system with a commensurate interaction ($p=1/2$) has been investigated experimentally~\cite{Landig_16} and numerically~\cite{Dogra_16,Sundar2016,Chen2016,Niederle2016,Flottat_17}, and a rich phase diagram, including a supersolid phase, has been explored. Here, we focus on a typical incommensurate case, e.g., $p = 1/\sqrt{2}$. To gain some insight, we first study the ground state in the limit $J=0$, where the problem becomes a classical spin glass problem whose ground state energy can be estimated by the classical simulated annealing method. The total ground-state energy $E_{\rm c}$ of the Hamiltonian (\ref{eq:hamilt}) in this  limit ($J = 0$) as a function of lattice length is plotted in Fig.~\ref{fig:mf_qmc_plots}(a), from which we find that, even though $E_{\rm c}$ dramatically oscillates with $L$, as a consequence of the incommensurate infinite long-range interaction, the lower enveloping line of the $E_c\textendash L$ curve linearly depends on $L$, indicating that in the thermodynamic limit, the average energy per site is independent of the system size. To determine the nature of this classical ground state, we introduce an order parameter to measure the incommensurate charge density wave (CDW) phase with period $1/p$, $m_c\equiv\frac{1}{L}\sqrt{\sum_{i,j}\left(n_i - \frac{1}{2}\right)\left(n_j - \frac{1}{2}\right)e^{\imath 2\pi p(i-j)}}$. As shown in Fig.~\ref{fig:mf_qmc_plots}(b), when $L\to\infty$, $m_c$ extrapolates to a finite value, indicating a long-range CDW correlation in the classical ground state. Now, we turn on the hopping in the above classical picture, which plays a role in inducing quantum fluctuations. To study the ground state of this model, we perform quantum Monte Carlo (QMC) simulations with a worm algorithm update~\cite{Prokofev_1998, Pollet_2007, Pollet_2009}, which is free from the sign problem since the frustration appears only in the diagonal (interacting) part of $\hat H$.

To determine the properties of the ground state, we calculate the superfluid density $\rho_s$ (computed using the winding number $W$~\footnote{In our world-line QMC simulation, each world-line is a closed curve defined in a torus; the winding number is the net number of times the world-lines wrap around the 1D system.} of the QMC simulations as $\rho_s \equiv (L/\beta)\langle W^2\rangle$) and perform the finite size scaling with the dynamical critical exponent $z = 1$, with the scaling relation between the space and imaginary time as $L/\beta = 1$. As shown in the classical case, the physical quantities in the ground state may dramatically oscillate with the lattice length, thus special attention needs to be taken for the finite size scaling. To derive the properties in the thermodynamic limit, we choose lattices whose length is close to the lower enveloping line of the $E_c\textendash L$ curve. The superfluid density $\rho_s$ as a function of interaction strength $V$ is plotted in Fig.~\ref{fig:mf_qmc_plots}(c), from which we find that for sufficiently large $V$, $\rho_s$ vanishes, indicating the breakdown of the quasi-superfluidity. In the inset of Fig.~\ref{fig:mf_qmc_plots}(c), we obtain a scaling invariant point at $V_c = 9.9 \pm 0.1J$, which indicates a continuous phase transition with dynamical critical exponent $z = 1$, separating a Luttinger liquid and a strongly interacting phase.

Now, we analyze the nature of the strongly interacting regime, which should adiabatically connect to the classical limit $J=0$. In Fig.~\ref{fig:mf_qmc_plots}(d), we plot the equal-time single-particle correlation function $G(x) = \langle \hat a^\dagger_i \hat a_{i+x}\rangle$, which decays exponentially in distance. Inspired by the results in the classical limit, we can introduce the incommensurate CDW order parameter $m_c$ and decouple the interaction under mean-field approximation in the thermodynamic limit and the Hamiltonian can be rewritten as
\begin{equation}
\nonumber H_{\rm MF} = -J\sum_i (a^\dagger_i a_{i+1} + h.c.)-2 V m_c \cos(2\pi p i)n_i + V m_c^2,
\end{equation}
with $p = 1/\sqrt{2}$. Hence, this mean-field Hamiltonian is equivalent to the Aubry-Andr\'e model (See Appendix~\ref{app:mf}), which is known to have a phase transition when increasing the strength of the incommensurate external potential~\cite{Aubry_80}. However, in our case, the translational symmetry breaks spontaneously instead of explicitly.

\begin{figure}[!t] 
  \includegraphics[width=0.95\columnwidth]{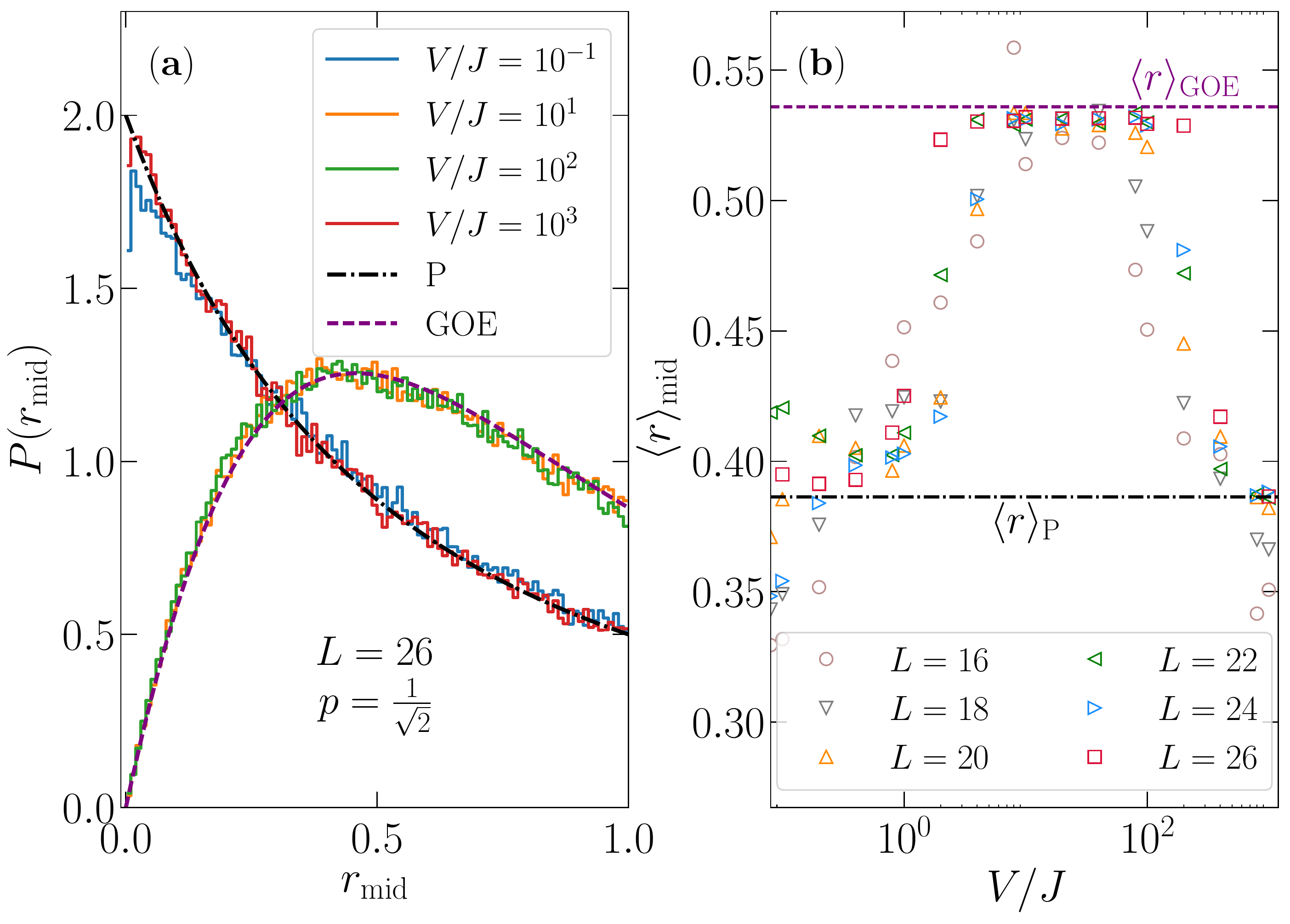}
 \vspace{-0.2cm}
 \caption{(Color online) Statistics of adjacent gaps probing ergodicity/non-ergodicity. (a) The distribution of the ratio of adjacent gaps in the central half of the spectrum ($L=26$) and different magnitudes of the interactions. The dashed (dashed-dotted) line depicts the analytical expression of an ergodic (nonergodic) quantum system described by  GOE (Poisson) statistics. (b) The average value of this quantity as $V$ is varied: for small and large values, nonergodic behavior sets in, while for intermediate interactions, quantum chaos prediction is obtained. To decrease statistical fluctuations, all values are averaged between the equivalent parity sectors of the system's symmetries (see text). Dashed lines describe the ergodic (non-ergodic) regimes with $\langle r\rangle \approx 0.5359$ (0.3863)~\cite{Atas_Bogomolny_13}.}
 \label{fig:ave_roag_vs_V_p1_sqrt_2_real_sectors}
\end{figure}

Glassy behavior in the absence of disorder also appears in other contexts of systems with long-range interactions either in a lattice~\cite{Angelone_2016} or in the continuous~\cite{Mendez_2017}.
\section{Finite energy densities: Statistical properties}
It is an open question whether localization present in the ground state of a system could be connected to the localization phenomenon in highly excited states. In our model, specifically, we aim to verify if the previously described glasslike phase also survives at finite energy densities. We begin with a diagnostic of quantum chaotic behavior~\cite{srednicki1994chaos, deutsch1991quantum, srednicki_99}, which has been widely used as a way to probe ergodicity in quantum systems. This can be quantified by the presence or not of energy level repulsion~\cite{Bohigas_Giannoni_84, Brody_Flores_81, Haake_91}, which in turn, can be measured via the ratio of adjacent gaps in the spectrum~\cite{Oganesyan_Huse_07, Atas_Bogomolny_13}, $r_\alpha \equiv \min\left(\delta_{\alpha+1},\delta_{\alpha}\right)/\max\left(\delta_{\alpha+1},\delta_{\alpha}\right)$, and $\delta_\alpha = E_{\alpha}-E_{\alpha-1}$ are gaps in between consecutive energy levels in the ordered list of eigenenergies $\{E_\alpha\}$ of the Hamiltonian. We show in Fig.~\ref{fig:ave_roag_vs_V_p1_sqrt_2_real_sectors}(a) the distribution of $r$ in the central half of the spectrum\footnote{Despite having long-range interactions, the Hamiltonian still possesses at most two-body terms. Thus, one only expects to see full level repulsion for eigenenergies away from the ends of the spectrum~\cite{Santos_Rigol_10a, Santos_Rigol_10b}}, averaged among all the real sectors of the Hamiltonian, obtained by using a basis that encodes translation, particle-hole, and inversion symmetries (See Appendix~\ref{app:symmetries} for a description of the symmetry resolving), for the largest system size we study, $L=26$ (Hilbert space dimension ${\cal D}\sim 10^5$). In the limit $V/J\ll1$, the system is close to the integrable regime and level repulsion is absent: the level spacings are completely uncorrelated and a Poisson distribution is obtained~\cite{Atas_Bogomolny_13}. On the other hand, for increasing interactions, the distribution becomes equivalent to the ones of symmetric random matrices belonging to a Gaussian orthogonal ensemble (GOE)~\cite{Atas_Bogomolny_13}, which is characteristic of thermalizing quantum systems. Nevertheless, when approaching a regime where the interaction strength is much larger than the hopping scale, a Poisson distribution for $r$ is once again recovered. This suggests that localization, described by the breakdown of ergodicity, at \textit{infinite temperatures} is obtained in a manner similar to the scenario of MBL for large enough quenched disorder, but here in a translation-invariant system.

The Hamiltonian (\ref{eq:hamilt}) is nonintegrable for any finite value of the interactions so one would expect the predictions of the eigenstate thermalization hypothesis (ETH)~\cite{srednicki1994chaos, deutsch1991quantum, srednicki_99, Rigol_Dunjko_08} to be numerically obtained, apart from finite size effects \footnote{To the best of our knowledge, the Hamiltonian is still non-integrable even when approaching the atomic limit ($V/J\to\infty$) for the case of incommensurate long-range interactions.}. To rule these out, we check in Fig.~\ref{fig:ave_roag_vs_V_p1_sqrt_2_real_sectors}(b) the average value of these distributions for different system lattice sizes and interactions. The intermediate ergodic regime ($V/J\sim10^1-10^2$) is robust and level repulsion, characterized by $r\sim r_{\rm GOE}$, is absent for $V/J\gtrsim 10^2$. As for the ground state, a proper finite-size scaling is rather elusive~\footnote{A proper finite-size scaling is even more elusive here in the case of finite energy densities given that the Hilbert spaces grow exponentially with the system size, which poses a challenge to exact diagonalization studies.}.

\section{Dynamical localization}

\begin{figure}[!t] 
  \includegraphics[width=1\columnwidth]{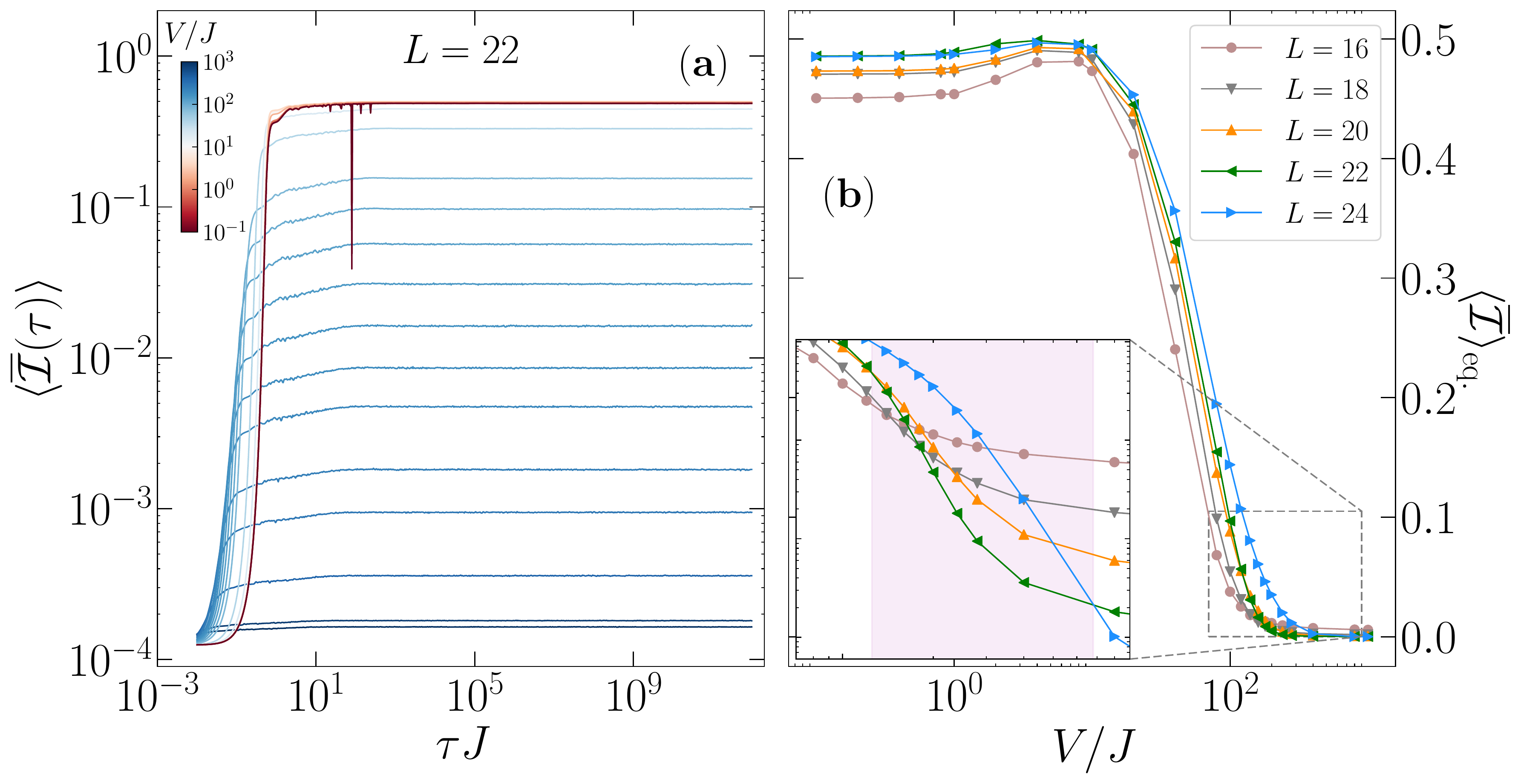}
 \vspace{-0.7cm}
 \caption{(Color online) (a) Time dependence of the dynamical IPR for a lattice with $L=22$ averaged over all equivalent real parity sectors, and different initial states, for several interaction strengths given by the color code. Panel (b) shows its equilibrium value in the long-time limit $\langle \overline{\cal I}\rangle_{\rm eq.}\equiv \langle \overline{\cal I}(\tau\to\infty)\rangle$ as a function of $V/J$, comparing the size dependence. The inset highlights the crossing region marking the beginning of the MBL-like regime; the shaded area represents the confidence interval for the critical interaction.}
 \label{fig:IPR_vs_t_and_IPR_eq_vs_V}
\end{figure}

A clearer picture of whether a transition from thermalization to MBL-like behavior takes place, at large interaction values, can be quantified in the dynamical properties of the system. We start by investigating the dynamical inverse participation ratio~\cite{van_Horssen_15, Barbiero_Menotti_15} (IPR), which is a measure of localization in the many-body basis, ${\cal I}_{\vec j}(\tau) = {\cal D}^{-1}[\sum_{\vec i} |\langle \psi_{\vec j}(\tau)|{\vec i}\rangle|^4]^{-1}$, obtained after time evolving an initial state $|{\vec j}\rangle$ ($|\psi_{\vec j}(\tau)\rangle = e^{-\imath\hat H \tau}|{\vec j}\rangle$) in the fully symmetric basis of $\hat H$. To reduce statistical effects, we average over \textit{all} initial states in each sector as well as the results from different (real) symmetric ones, obtaining $\langle\overline{\cal I}(\tau)\rangle$. At long times, a time-evolved state is localized [delocalized] in the basis if the dynamical IPR is proportional to $O({\cal D}^{-1})$ $[O(1)]$. Figure~\ref{fig:IPR_vs_t_and_IPR_eq_vs_V}(a) shows that with increasing interactions, the equilibrium value of the dynamical IPR is substantially reduced and the system may fully retain information about its initial preparation at large time-scales. To account for finite size effects, we compare the equilibrium value $\langle\overline{\cal I}(\tau)\rangle_{\rm eq.}$ of the dynamical IPR for different lattices in Fig.~\ref{fig:IPR_vs_t_and_IPR_eq_vs_V} (b).  A crossing of the curves is obtained for values of $V/J$ in the range $\sim130-680$ [see inset in Fig.~\ref{fig:IPR_vs_t_and_IPR_eq_vs_V}(b)], after which one expects that localization survives in the thermodynamic limit, defining the critical value of interactions where the MBL-like behavior takes place and thermalization no longer holds.

\section{Initial state inhomogeneities}
\begin{figure}[!tb] 
  \includegraphics[width=1\columnwidth]{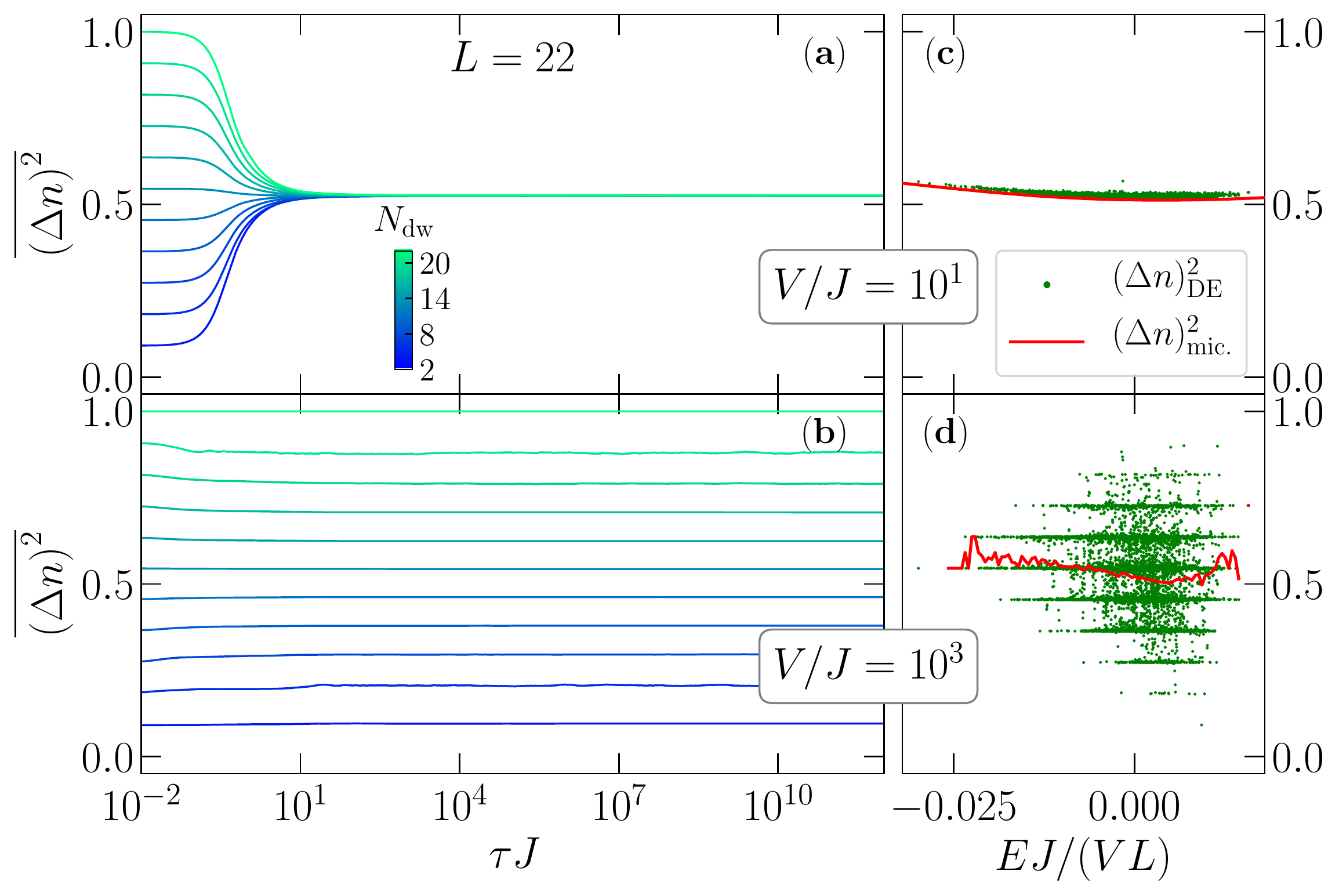}
 \vspace{-0.6cm}
 \caption{(Color online) Time evolution of the integrated charge inhomogeneity $\overline{(\Delta n)^2}$ averaged over states with similar $N_{\rm dw}$ in ergodic [nonergodic] regimes in panel (a)[panel (b)] with $V/J = 10^1$ [$V/J = 10^3$]. Panels (c) and (d) are comparisons of the DE ensemble prediction and the microcanonical result for the corresponding values of interactions.}
 \label{fig:ave_int_delta_n_sqd_t_time_evolution_k0_sector}
\end{figure}

We further characterize the interaction-induced quantum glass transition by noting how an inhomogeneity in the particle distribution in real space for the initial state persists after the unitary time-evolution~\cite{Carleo_12, Schiulaz_15, van_Horssen_15}. This has the advantage of being relevant to experiments in optical lattices that can probe site-resolved densities. We define the charge inhomogeneity operator $(\Delta \hat n)^2 \equiv (1/L)\sum_i(\hat n_{i+1} -\hat n_{i})^2$, whose expectation values within the initial symmetric Fock states are in the range [$\frac{2}{L}, \frac{4}{L}, \dotsc, 1$], quantifying the number of domain walls ($N_{\rm dw}$) present in these states. Figures~\ref{fig:ave_int_delta_n_sqd_t_time_evolution_k0_sector} (a) and \ref{fig:ave_int_delta_n_sqd_t_time_evolution_k0_sector}(b) display the time-integrated charge inhomogeneity, $\overline{(\Delta n)^2}(\tau) = (1/\tau)\int_0^\tau(\Delta n)^2(\tau^\prime)d\tau^\prime$, in the thermal [Fig.~\ref{fig:ave_int_delta_n_sqd_t_time_evolution_k0_sector}(a) - $V/J=10^1$] and nonergodic regimes [Fig.~\ref{fig:ave_int_delta_n_sqd_t_time_evolution_k0_sector}(b) - $V/J=10^3$], averaged over \textit{all} the initial states with equivalent $N_{\rm dw}$. The differences are clear: in the former, the time evolution results in a featureless state irrespective of the value of $(\Delta n)^2(0)$, while in the latter the charge inhomogeneity of the initial state is preserved for arbitrarily long times. Appendix~\ref{app:other_p} shows that this is intrinsically related to the incommensurability of the interactions with respect to the lattice spacings by checking other values of periodicities $p$. In contrast, if one considers commensurate interactions as $p=1/2$, information about the initial preparation is eventually lost as shown in Appendix~\ref{app:commens}. Figures ~\ref{fig:ave_int_delta_n_sqd_t_time_evolution_k0_sector}(c) and \ref{fig:ave_int_delta_n_sqd_t_time_evolution_k0_sector}(d) compare the infinite-time average, given by the diagonal ensemble (DE), for all the initial symmetric Fock states with the corresponding microcanonical prediction~\cite{srednicki_99, Rigol_Dunjko_08} showing the breakdown of thermalization in the strongly interacting regime. Appendix~\ref{app:L26} displays similar results for a larger lattice size, $L=26$.

\section{Many-body density of states and ETH}
\begin{figure}[!tb] 
  \includegraphics[width=1\columnwidth]{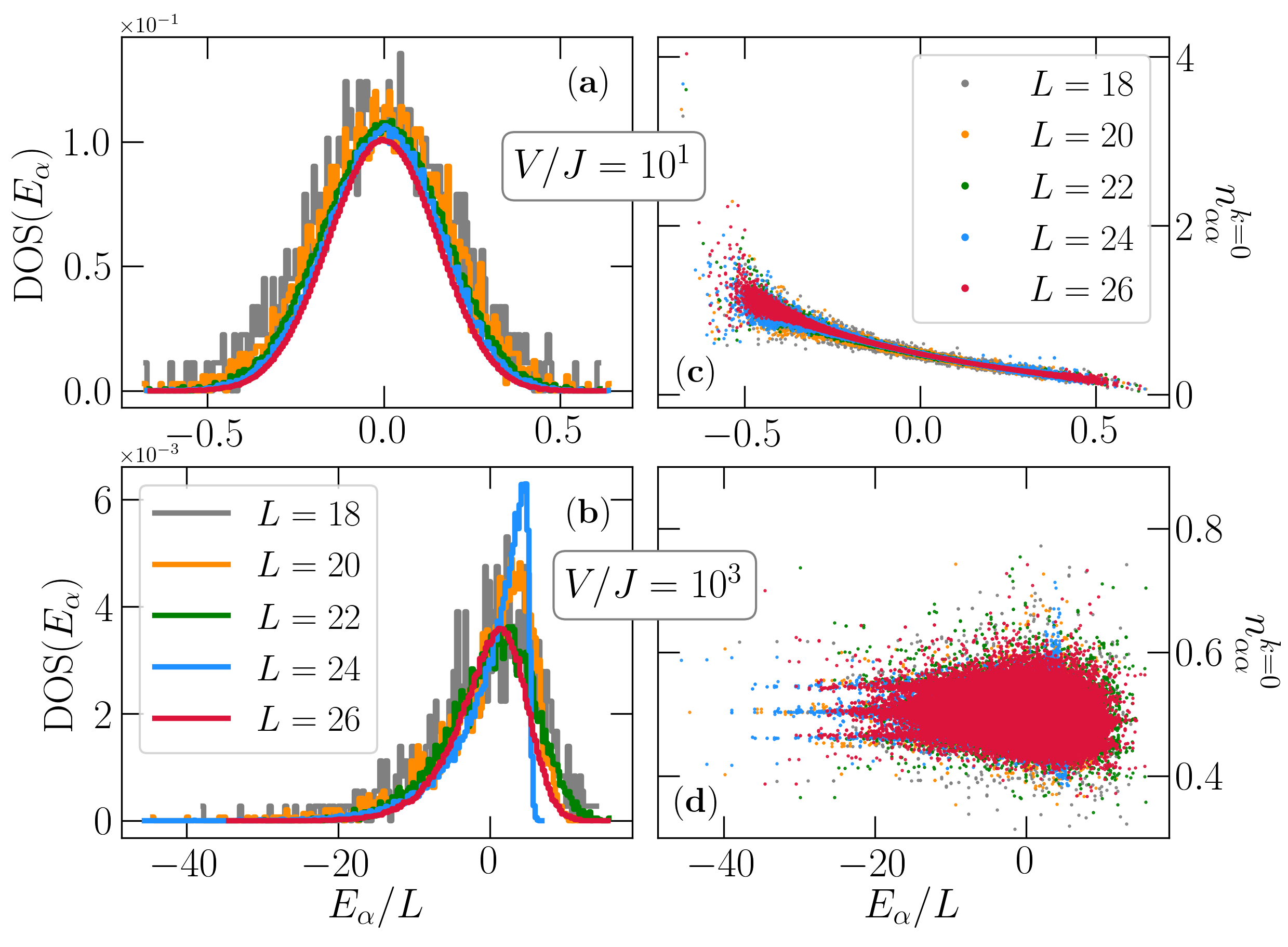}
 \vspace{-0.6cm}
 \caption{(Color online) Normalized density of states [panels (a) and (b)] and eigenstate expectation values of the zero-momentum occupation $n^{k=0}_{\alpha\alpha}$ [panels (c) and (d)] for $V/J=10$ [panels (a) and (c)] and $V/J=10^3$ in [panels (b) and (d)]. The clear ETH-like behavior for $n^{k=0}_{\alpha\alpha}$ in panel (c) is lost once interactions are further strengthened (d).}
 \label{fig:dos_nk0_alpha_alpha}
\end{figure}
When investigating a translation-invariant MBL system, one needs to be careful with the possible formation of ``mini-bands'' in the spectrum of finite systems~\cite{Papic_15}. These are typical in proposals of translation-invariant MBL consisting of two types of particles (light and heavy)~\cite{Yao_Laumman_16, Papic_15} with very different energy scales; they reduce the effective Hilbert space and do not persist in the thermodynamic limit, resulting in pseudo-MBL behavior that is essentially a finite size effect. To test this scenario, Fig.~\ref{fig:dos_nk0_alpha_alpha} (a) and \ref{fig:dos_nk0_alpha_alpha}(b) show the normalized density of states (DOS) with increasing magnitudes of interactions. Albeit some rather large finite-size effects affect the distribution shape (see, e.g., $L=24$ results and Appendix~\ref{app:entropy} concerning the monotonicity of thermodynamic quantities), the DOS does not display a separation into small bands for the case of a real even (under all parity symmetries) sector of the Hamiltonian, even at the largest interaction $V/J = 10^3$, which already displays an MBL-like behavior for other quantities. Besides, it is clear that in this strong interaction regime,  thermalization, as prescribed by the  ETH, fails with the support of the eigenstate expectation values of a few-body operator, like the zero momentum occupation [$\hat n^{k=0}=(1/L)\sum_{i,j}\hat a_{i}^\dagger \hat a_{j}$] displayed in Figs.~\ref{fig:dos_nk0_alpha_alpha} (c) and \ref{fig:dos_nk0_alpha_alpha}(d), not decreasing with system size.

\section{Summary}
Using large-scale numerical calculations, ranging from classical annealing to QMC to exact diagonalizations, we study a quantum glass transition of a system composed of hard-core bosons subjected to quasiperiodic long-range interactions. This translation-invariant system displays a transition to localized behavior at sufficiently large interaction magnitudes not only at the groundstate but also at finite energy densities. In the latter, we associate it to an MBL-like regime where the breakdown of thermalization and initial state memory survives at exponentially large time scales, even in the absence of any disorder. Although the infinite time-average for initial states with typical values of $N_{\rm dw}$ may approach the thermal prediction in the large interaction limit, the number of (initial) states that result in a discrepancy is not necessarily rare, suggesting that initial memory survives indefinitely, even at infinite temperatures (see Appendix~\ref{app:L26}), unlike in the scenario of translation-invariant MBL proposals of ``light'' and ``heavy'' particles. Nevertheless, the recent emulation of long-range interactions in optical lattices embedded in high-finesse cavity experiments, emulating the physics described here, may settle this issue by probing whether the so far numerically observed out-of-equilibrium localization in systems without quenched disorder is a phenomenon that is not related to finite size effects. 

\begin{acknowledgments}
R.M. is financially supported by the National Natural Science Foundation of China (NSFC) (Grants No. U1530401, No. 11674021, and No. 11650110441).  ZC is  supported by the National Key Research and Development Program of China (Grant No. 2016YFA0302001), the National Natural Science Foundation of China under Grant No.11674221, and the Shanghai Rising-Star Program. R.M. acknowledges enlightening discussions with Deepak Iyer and Marcos Rigol. Z.C. thanks L.~Pollet for valuable discussions. The computations were performed on the Tianhe-2JK at the Beijing Computational Science Research Center (CSRC). Z.C. acknowledges the support from the Center for High Performance Computing of Shanghai Jiao Tong University.\\
\end{acknowledgments}

\appendix

\section{Mean-field analysis}
\label{app:mf}
\begin{figure}[!] 
  \includegraphics[width=0.9\columnwidth]{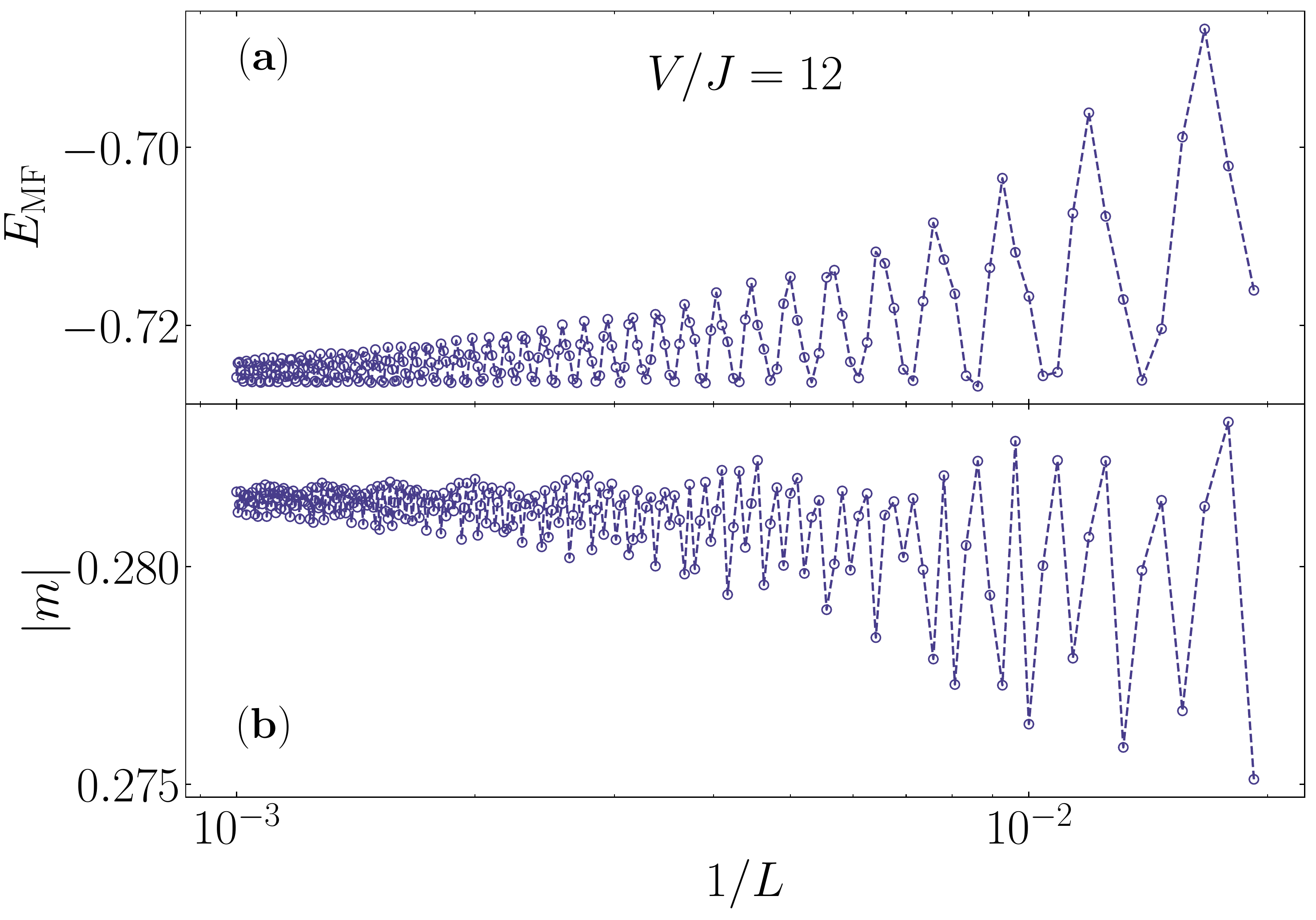}
 \caption{(Color online) (a) System size dependence of the ground-state energy per site obtained within the mean-field approximation and (b) the CDW order parameter obtained from the mean-field for an interaction strength $V/J=12$ with incommensurate interactions whose inverse period is $p=1/\sqrt{2}$.}
 \label{fig:mean_field_supp}
\end{figure}

We explore here in more details the mean field version of the Hamiltonian that is valid in the strongly interacting regime. The original Hamiltonian can be rewritten as
\begin{eqnarray} 
\nonumber &\hat H&=-J\sum_i (\hat a_i^\dagger \hat a_{i+1}+h.c)-\frac{V}{2L}\{[\sum_i e^{\imath2\pi pi}(\hat n_i-\frac 12)] \\
&\cdot& [\sum_j e^{-\imath2\pi pj}(\hat n_j-\frac 12)]+h.c\}. \label{eq:Hamil}
\end{eqnarray}
By introducing the mean-field charge-density-wave order parameter: $m=\frac 1L \sum_i  e^{-\imath2\pi pi}\left(\langle \hat n_i\rangle-\frac 12\right)$, the Hamiltonian in Eq.(\ref{eq:Hamil}) can be decoupled as:
\begin{eqnarray}
\nonumber \hat H&=&\sum_i -J(\hat a_i^\dag \hat a_{i+1}+h.c)\\
&-&\sum_i \{V [m e^{-\imath2\pi pi}(\hat n_i-\frac 12)+h.c]+\frac V2|m|^2\}.\label{eq:MFham}
\end{eqnarray}
The translational symmetry has been broken explicitly in the mean-field Hamiltonian~(\ref{eq:MFham}), where the mean-field order parameter $m$ can be solved using the standard  self-consistent method in real-space. In Fig.~\ref{fig:mean_field_supp} (a) and Fig.~\ref{fig:mean_field_supp} (b), we plot the result of the ground-state energy and the mean-field order parameter $|m|$ as a function of the system size $L$ for a fixed $V/J=12$, which according with the results from the QMC, is already in the strongly correlated regime. We note that the fluctuations markedly decrease for larger system sizes properly defining the thermodynamic limit. This behavior is also expected in classical case ($J=0$), originally presented in Figs.~1 (a) and (b) in the main text, provided one is able to reach larger system sizes.

\section{Symmetry sector resolving}
\label{app:symmetries}
To probe aspects related to quantum chaotic behavior (and the lack of thereof for large enough interactions) encoded in the level repulsion, or to compute infinite time-averages (diagonal ensemble averages), one most ensure that all ``trivial'' degeneracies are resolved. These are in general related to the symmetries of the Hamiltonian. In our model, these are three: translation ($\hat T_{x}$), and parities under particle-hole ($\hat P_{\rm phs}$) and reflection ($\hat P_{I}$). We construct a basis where the states are simultaneously eigenstates of these three symmetry operations, which reduces the total Hilbert space ${\cal D}={{L}\choose{L/2}}$ to smaller sectors ${\cal D^\prime}\approx {\cal D} / (L\cdot 2\cdot 2)$. We restrict ourselves to real sectors (with corresponding momentum quantum numbers $k=0$ or $\pi$) and, to reduce statistical fluctuations, we average over the 8 equivalent sectors after the parity operations in Figs.~2 and 3. Whereas in Figs.~4 and 5, we restrict to the even sectors under $\hat P_{\rm phs}$ and $\hat P_{I}$ for the zero momentum translation quantum number.

\section{Infinite time average and the initial memory conservation}
\label{app:L26}
\begin{figure}[t] 
  \includegraphics[width=1\columnwidth]{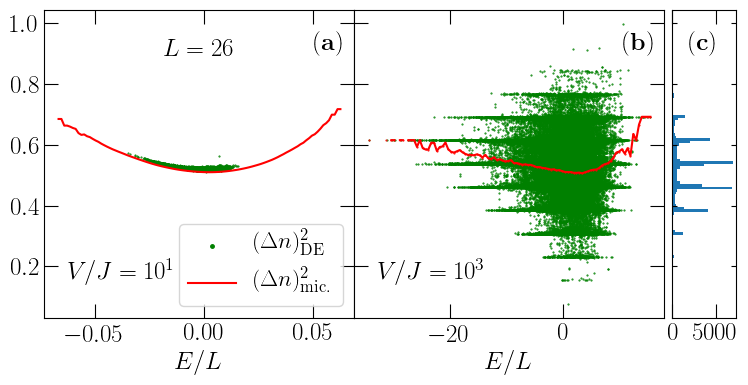}
 \vspace{-0.6cm}
 \caption{(Color online) Comparisons of the infinite time-average (diagonal ensemble prediction) of the charge inhomogeneity $(\Delta \hat n)^2 \equiv (1/L)\sum_i(\hat n_{i+1} -\hat n_{i})^2$ and the microcanonical (thermal) results $[(\Delta \hat n)^2)_{\rm mic.} = \sum_\alpha |\langle \alpha | \psi(0)\rangle |^2 (\Delta \hat n)^2_{\alpha\alpha}]$ in the ergodic (a) and non-ergodic (b) regimes for a lattice with 26 sites. Panel (c) depicts the histogram of the average charge inhomogeneities when $\tau\to\infty$ at \textit{infinite temperatures} (see text).}
 \label{fig:DE_mic_L26}
\end{figure}

The initial memory preservation is a characteristic of many-body localized systems. In Fig.~4 (main text), we show that, in the strongly interacting (non-ergodic) regime, a discrepancy is clear between the infinite time average of the charge inhomogeneities, encoded in the initial (symmetric) Fock states, and the values of a thermal ensemble, signaling the thermalization breakdown. Here, we repeat this analysis in Fig.~\ref{fig:DE_mic_L26} for the largest system size we study, presenting qualitatively similar results. The states for which the discrepancy is larger, are the ones that have a number of domain walls ($N_{\rm dw}$) far from the typical value, which displays infinite time predictions closer to the thermal result. We argue here that although the number of those states is small, they still represent a substantial fraction of the total number of initial Fock states. This can be seen by the histogram in panel (c) that counts the values of the diagonal ensemble results of $(\Delta \hat n)^2$ for initial states whose energies $E_0$ ($= \langle\psi(0)|\hat H|\psi(0)\rangle$) are within the central half of the system's eigenenergies. In the thermodynamic limit, the peak-structure will become smoother but with large tails that will manifest the lack of thermalization.

\section{Entropy at infinite temperatures}
\label{app:entropy}
\begin{figure}[t] 
 \includegraphics[width=0.75\columnwidth]{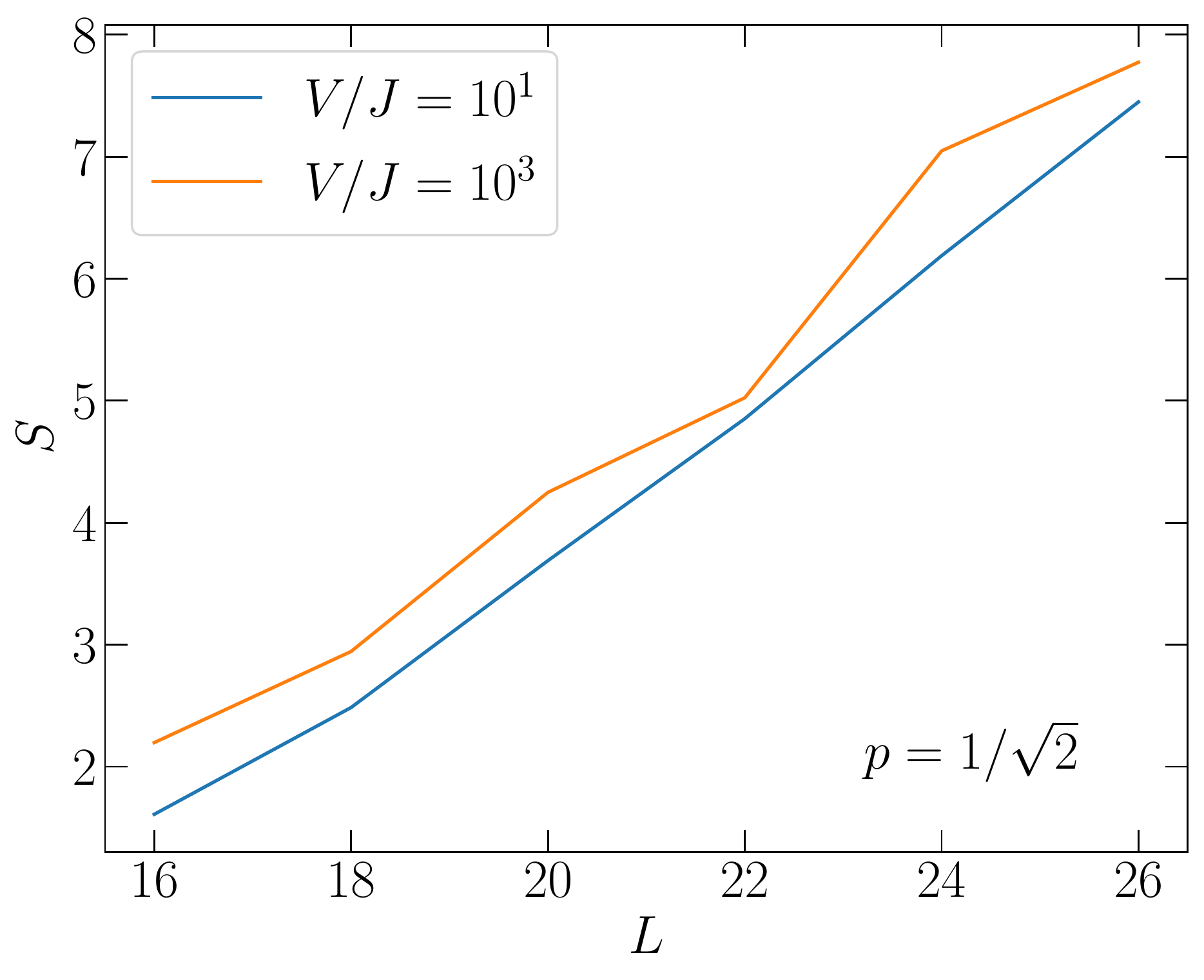}
 \vspace{-0.6cm}
 \caption{(Color online) Thernodynamic entropy at infinite temperatures vs. the lattice size in the ergodic ($V/J = 10^1$) and non-ergodic regimes ($V/J=10^3$) for a subsector of the Hamiltonian with zero total momentum and with even parities under particle-hole and reflection symmetries.}
 \label{fig:entropy_nx0}
\end{figure}

One concern that may arise with the manifest large finite-size effects appearing in the strongy correlated regimes, refers to the monotonicity of thermodynamic quantities. For example, the thermodynamic entropy $S$ at energy $E$ is related to the density of states via $e^{S(E)} = E\sum_{\alpha} \delta(E-E_\alpha)$. At infinite temperatures, one can write down the thermodynamic entropy as the logarithm of the density of states at its maximum value. Figure~\ref{fig:entropy_nx0} displays the system size dependence of this quantity for two values of interactions, $V/J = 10^1$ and $10^3$, in the ergodic and non-ergodic regimes, respectively. In the latter, the fluctuations are a direct manifestation of the finite-size effects but still show how the entropy is a momotonic function of the system size.

\section{Incommensurate case: other periodicities}
\label{app:other_p}
\begin{figure*}[!] 
  \includegraphics[width=0.75\textwidth]{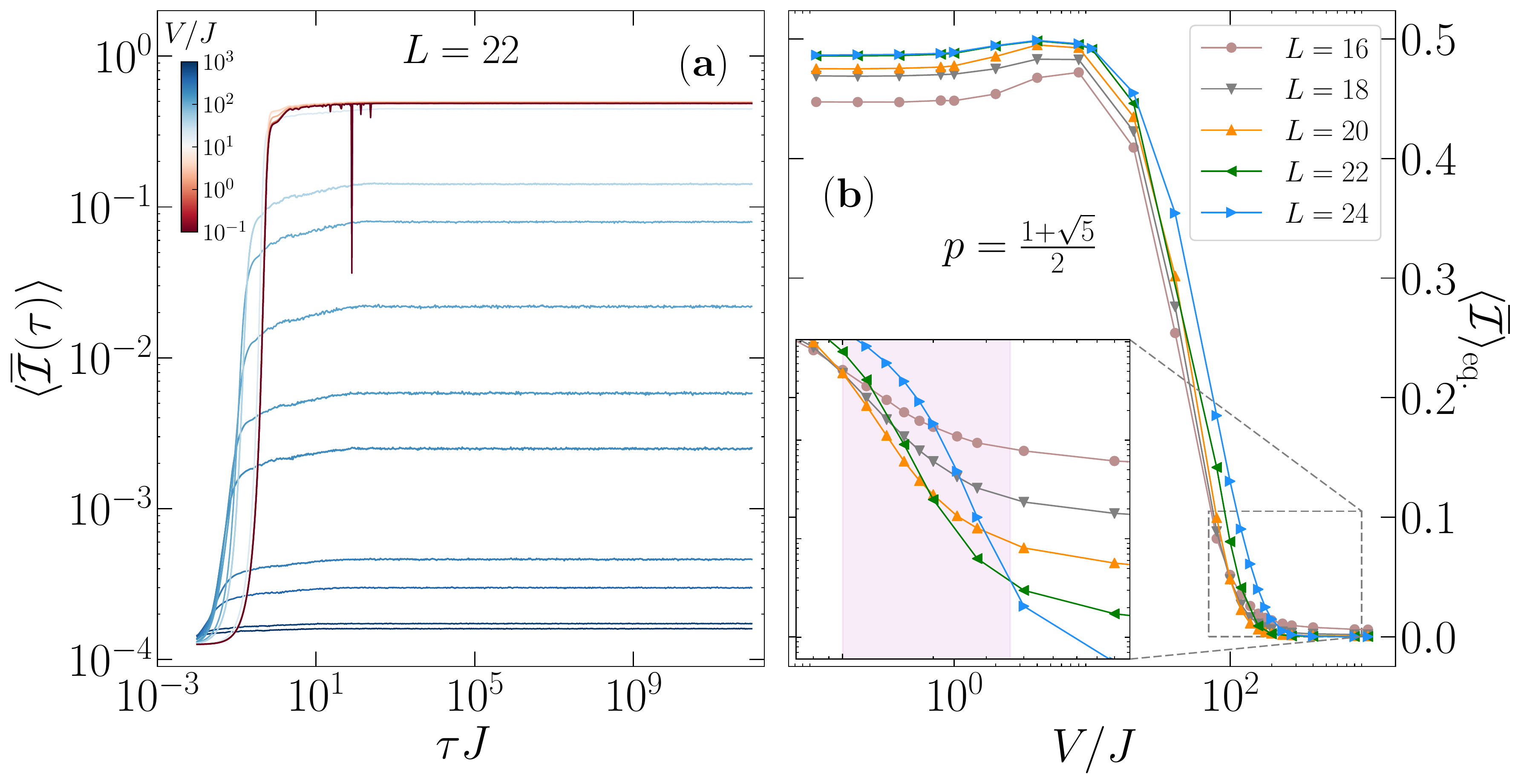}
 \vspace{-0.6cm}
 \caption{(Color online) Same as in Fig.~3. The crossing region, given by the shaded area in the inset of (b), depicts the confidence interval of the localization transition given the system sizes available: $V_c/J = 100-360$. This is consistent with the other periodicity $p=1/\sqrt{2}$ though the interval is reduced due to smaller finite-size effects.}
 \label{fig:IPR_vs_t_and_IPR_eq_vs_V_other_periodicity}
\end{figure*}
We have established that the system displays localizaton at infinite temperatures for the case of long-range quasi-periodic interactions with period $T=\sqrt{2}$. Nevertheless, finite size effects are strongly manifest in a variety of quantities when determining the boundary between the ergodic regime and the many-body localized one. One of the quantities that is largely affected is the dynamical inverse participation ratio (Fig.~3 in the main text), where the results for system size $L=24$ fail to reach a crossing point in $V/J$ compatible with other lattice sizes. That this size presents pathological behavior can also be inferred from the non-gaussian behavior of the many-body density of states in the strongly interacting limit (Fig.~5(b) in the main text), in contrast to other values of $L$. One of the reasons behind this is that the period of interactions is \textit{almost} commensurate with the lattice size for $L=24$ when $p=1/\sqrt{2}$.

Different approaches may be used to mitigate finite size effects. We choose a simple one in which we see how robust is the transition point under variations of the incommensurate periodicity of the interactions. We select a new period $T=1/p$ with $p$ given by the golden ratio $\frac{1+\sqrt{5}}{2}$ and report in Fig.~\ref{fig:IPR_vs_t_and_IPR_eq_vs_V_other_periodicity} the same as in Fig.~3 of the main text but for this different periodicity. We note that for the system sizes studied for this quantity (even number of lattice sites from $L=16$ to 24) the crossing point (around $V_c/J = 100-360$) is consistent with the previous estimation from $p=1/\sqrt{2}$.

\section{The commensurate case: $p=1/2$}
\label{app:commens}
The case of interactions whose period $T=1/p$ is twice the lattice constant corresponds to the realistic situation in the experiment described in Ref.~\cite{Landig_16}. In this scenario, all the lattice sites can be classified into even and odd groups and the interaction between the two particles in the different(same) group are repulsive (attractive) with the same interaction strength. The interaction can be rewritten as: $\hat H_{\rm int} = -\frac{V}{L}(\hat N_{e} - \hat N_{o} )^2$ where $\hat N_{e}(\hat N_{o})$ is the total particle number operator in the even (odd) sublattice. Notably, the interaction favors the formation of a charge-density-wave (CDW) state with a non-zero order parameter: $m = \langle \hat N_{e} - \hat N_{o}\rangle⟩/L$. A simple mean-field treatment of the Hamiltonian suggests that even an infinitesimal $V$ will open a gap and induce a CDW insulating phase with spontaneously translational symmetry breaking. 

\begin{figure}[t] 
  \includegraphics[width=1\columnwidth]{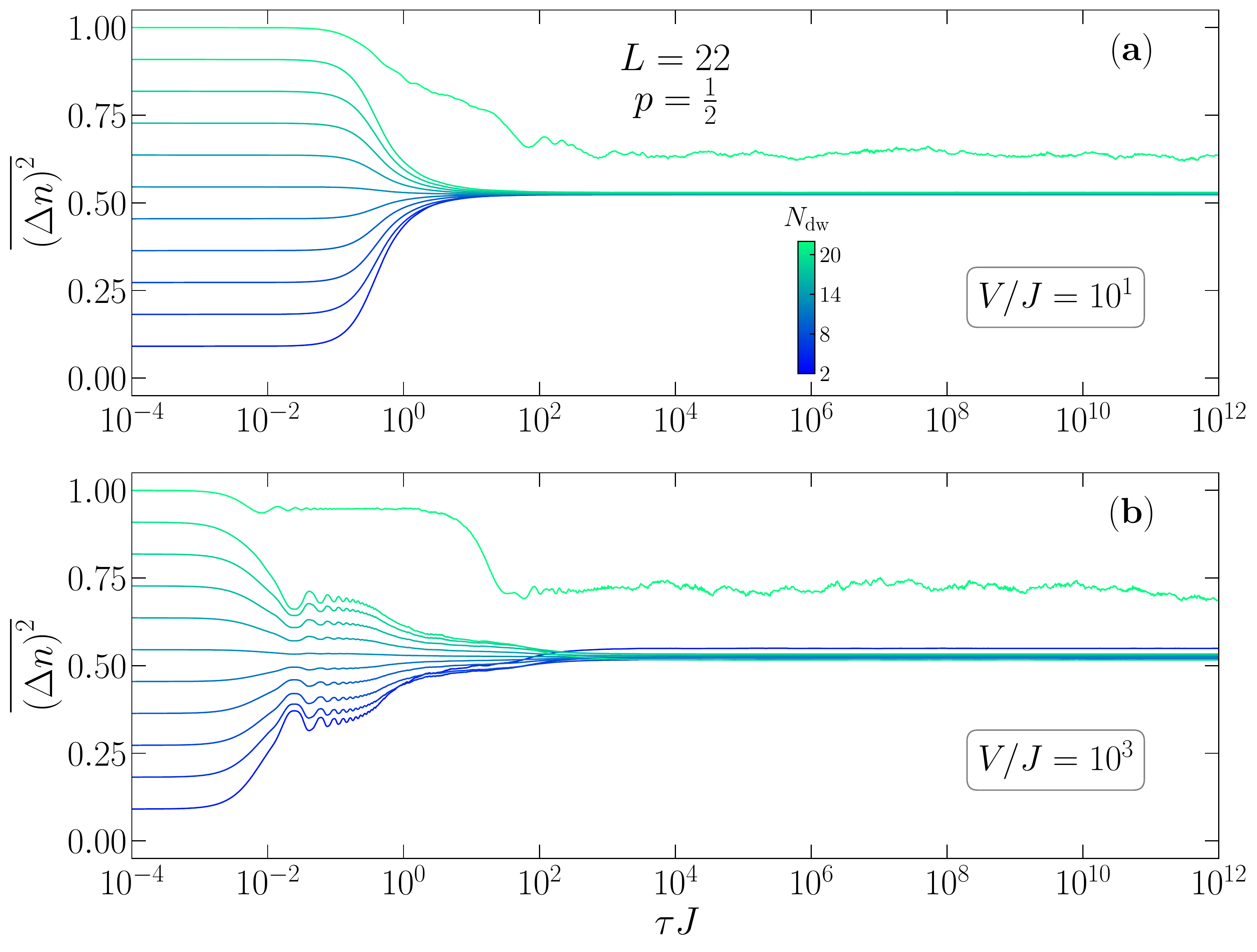}
 \vspace{-0.6cm}
 \caption{(Color online) Similar to Figs.~4(a) and 4(b) in the main text but for the case of interactions whose period is commensurate with the lattice spacing ($p=1/2$). The time-dependence of the integrated  charge inhomogeneity is reported for interactions $V/J = 10^1$[$10^3$] in (a)[(b)].}
 \label{fig:ave_int_delta_n_sqd_t_time_evolution_k0_sector_commensurate}
\end{figure}

In the out-of-equilibrium situation, to highlight that this model does \textit{not} give similar conclusions as to the incommensurate case $p=1/\sqrt{2}$ studied in the main text, we show in Fig.~\ref{fig:ave_int_delta_n_sqd_t_time_evolution_k0_sector_commensurate} the quantity which can be used experimentally to probe localization, the degree of charge inhomogeneity of time-evolved initial states.

For similar values of interactions which are in the ergodic and non-ergodic regimes for $p=1/\sqrt{2}$, $V/J=10^1$ and $10^3$, respectively, we see that the initial charge inhomogeneity vanishes in both cases, resulting in a featureless state for large enough time-scales in contrast to the localization observed for large interaction values in the case of incommensurate interactions. As in the main text, we average the time-integrated charge inhomogeneity, $\overline{(\Delta n)^2}(\tau)$, for states with equivalent number of domain walls that quantifies the degree of inhomogeneity. The notable exception is the initial state with maximum number of domain walls $[\overline{(\Delta n)^2}(0)=1]$, corresponding to a CDW initial state \textendash~in fact, a symmetric version incorporating all the symmetries of the basis. This state approaches the actual ground-state of the Hamiltonian in the large $V$ limit and possess total energy that is largely gapped from the bulk of the spectrum. This results in a lack of hybridization with other states with different degrees of charge-inhomogeneity, then partially preserving information of the initial state. Yet it does not constitute though a localization feature at \textit{infinite temperatures}, characteristic of many-body localization. 

We note as well that in the irreducible sectors of the Hamiltonian \textendash~ after applying translation, inversion and particle-hole symmetries \textendash~ the spectrum does not display level repulsion in any range of the interaction magnitude. That is suggestive that the Hamiltonian might display integrability for commensurate interactions as, e.g., for $p=1/2$. Thus, this precludes a scenario of breakdown of ergodicity signaling the onset of translation-invariant many-body localization.

\bibliography{self_loc}

\end{document}